\documentclass[a4paper]{jpconf}
\usepackage{graphicx}
\begin{document}
\title{The multiplicative Hamiltonian and its hierarchy}

\author{Saksilpa Srisukson$^{a,1}$, Kittikun Surawuttinack$^{b,2}$, Sikarin Yoo-Kong$^{b,c,d,3}$}

\address{$^a$ Sarasas Ektra School 336/7 Sathupradit Rd., Bangpongpang, Yannawa , Bangkok 10120, Thailand.\\
$^b$ Theoretical and Computational Physics (TCP) Group, Department of Physics, Faculty of Science, King Mongkut's University of Technology Thonburi, Bangkok 10140, Thailand. \\
$^c$ Theoretical and Computational Science Centre (TaCs), Faculty of Science, King Mongkut's University of Technology Thonburi, Bangkok 10140, Thailand.\\
$^d$ Ratchaburi Campus, King Mongkut's University of Technology Thonburi, Ratchaburi, 70510, Thailand.}

\ead{$^{1}$saksilpa@gmail.com,$^{2}$brooklyn13306@hotmail.com,$^{3}$syookong@gmail.com}

\begin{abstract}
The multiplicative Hamiltonian flow on the phase space for a system with 1 degree of freedom was constituted from infinite hierarchy Hamiltonian flows. A new type of canonical transformation associated with the multiplicative Hamiltonian was found and called the $\lambda$-extended class of the standard canonical transformations.
\end{abstract}

\section{Introduction}
It was well known that Lagrangian processes the non-uniqueness property. This means that Lagrangian\footnote{The standard Lagrangian is in the additive form: $L=T-V$, where $T$ is the kinetic energy and $V$ is the potential energy} can be added by a constant, multiplied by a constant or added by the total derivative term leaving the same Euler-Lagrange equation. Recently, there has been found a new type of non-uniqueness of the Lagrangian called the multiplicative form of the Lagrangian for a system with 1 degree of freedom \cite{Su}. This new form of the Lagrangian comes with an extra-parameter namely $\lambda$ which can be considered as the parameter reducing the multiplicative Lagrangian to the additive Lagrangian. Furthermore, the multiplicative Lagrangian can also be considered as a generating function to produce an infinite hierarchy of additive Lagrangians, which produce the same equation of motion.\\\\
In this study, we focus on the properties of the multiplicative Hamiltonian. Since we knew there is the Legendre transformation connecting between Lagrangian and Hamiltonian, we will find the Legendre transformation for these multiplicative Lagrangian and Hamiltonian and its hierarchy. We also knew that Hamiltonian is a time generator to produce what we call the Hamiltonian flow on the phase space. We then would like to investigate this structure for the multiplicative Hamiltonian and its hierarchy. Moreover, working with the Hamiltonian setting on phase space, we can choose many sets of canonical coordinates through four types of canonical transformations (CTs). Then we will find new four types of CTs for the multiplicative Hamiltonian.

\section{The hierarchy}
\subsection{Legendre transformation}
The multiplicative Lagrangian and Hamiltonian have been recently discovered by Surawuttinack et al. [1], given by
\begin{equation}
L_{\lambda}(x,\dot{x}) = m\lambda^{2}\Big(e^{-\frac{\dot{x}^2}{2\lambda^2}} + \frac{\dot{x}}{\lambda^2} \int_0^{\dot{x}}e^{-\frac{v^2}{2\lambda^2}}dv \Big) e^{-\frac{V(x)}{m\lambda^2}}\; , \;
H_{\lambda}(p,x) = -m\lambda^{2}e^{-\frac{H_N}{m\lambda^2}}\;, 
\end{equation}
where $H_N=T+V$ is the standard additive Hamiltonian. The constant $\lambda$, which is in the unit of velocity, can be considered as a reduction parameter sending the multiplicative to additive form as follows; $\lim_{\lambda \to \infty} L_\lambda - m\lambda^2 = T - V$. The multiplicative Lagrangian and Hamiltonian can be rewritten in the form
\begin{equation}\label{Llam}
L_\lambda(x,\dot{x}) = \sum_{j=1}^{\infty}\frac{1}{j!}\Big(-\frac{1}{m\lambda^2}\Big)^{j-1}L_j + m\lambda^2 \; , \;
H_\lambda(p,x) = \sum_{j=1}^{\infty}\frac{1}{j!}\Big(-\frac{1}{m\lambda^2}\Big)^{j-1}H_j - m\lambda^2\;,
\end{equation}
 where 
\begin{equation}
L_j=\sum_{k=0}^j\Big[\frac{j!T^{j-k}V^k}{(j-k)!k!(2j - (2k+1))}\Big] \; , \;
H_j = (T + V)^j=H_N^j. 
\end{equation}
Lagrangians $L_j$ and Hamiltonians $H_j$ form the infinite hierarchies which they produce the same equation of motion,  $m\ddot{x} = -dV/dx$. We find a new momentum variable associated with multiplicative Lagrangian given by 
\begin{equation}
p_\lambda (p,x) = \frac{\partial L_\lambda}{\partial \dot{x}}= m \Big(\int_0^{\dot{x}}e^{-\frac{v^2}{2\lambda^2}}dv\Big) e^{-\frac{V(x)}{m\lambda^2}} = \sum_{j=1}^{\infty}\frac{1}{j!}\Big(-\frac{1}{m\lambda^2}\Big)^{j-1}p_j\;,
\end{equation}
where
\begin{equation}
p_j (p,x)=j!\Big[p_{j-1}V(x) + \frac{p^{2j-1}}{(j-1)!(2^{j-1})(2j-1)m^{j-1}}\Big];\;\;\;j\geq1\; \mbox{and} \; p=m\dot{x}.
\end{equation}
We find that the multiplicative Lagrangian and Hamiltonian are connected through the Legendre transformation $L_\lambda=p_\lambda\dot{x} - H_\lambda$ \cite{Su}. Substituting eq. (\ref{Llam}) into the Legendre transformation, we obtain
\begin{equation}\label{LG}
0 = \sum_{j=1}^{\infty}\frac{1}{j!}\Big(-\frac{1}{m\lambda^2}\Big)^{j-1}\Big[L_j - p_j\dot{x} + H_j\Big]\Rightarrow L_j = p_j\dot{x} - H_j\;,
\end{equation}
which results in an infinite hierarchy of Legendre transformations connecting between $L_j$ and $H_j$.

\subsection{Hamilton's equations}
From eq. (\ref{LG}), we consider the total derivative $dH_j(p,x) = d[p_j(p,x)\dot{x}] - dL_j(x,\dot{x})$ leading to
\begin{equation}\label{Hm}
dx\left(\frac{\partial H_j}{\partial x} +\frac{\partial p_j}{\partial p}\dot{p}\right) + dp\left(\frac{\partial H_j}{\partial p} - \dot{x}\frac{\partial p_j}{\partial p}\right) = 0.
\end{equation}
Equation (\ref{Hm}) holds if 
\begin{equation}
\frac{\partial H_j}{\partial x} = - \frac{\partial p_j}{\partial p}\dot{p} \: , \:
\frac{\partial H_j}{\partial p} = \frac{\partial p_j}{\partial p}\dot{x}\;,
\end{equation}
which are new forms of Hamilton's equations for each $H_j$ and they 
form the infinite hierarchy of the Hamilton's equations which again produce the same equation of motion. Obviously, for the case $j=1$, we can retrieve the standard Hamilton's equations since $p_1=p$.

\subsection{Multiplicative Hamiltonian flows on phase space}
We now define $\xi \equiv (p,x)$ and its Poisson bracket with $H_\lambda$ results in the $\lambda$-flow given by
\begin{equation}\label{F1}
\frac{d}{dt_\lambda}\xi =\Big\{\xi, H_\lambda\Big\} = \Big\{\xi,  \sum_{j=0}^{\infty}\frac{1}{j!}\Big(-\frac{1}{m\lambda^2}\Big)^{j-1}H_j \Big\}
\end{equation}
or
\begin{equation}\label{F2}
\frac{d}{dt_\lambda}\xi = \Big(\frac{\partial}{\partial t_1}+\frac{\partial}{\partial t_2}+...\Big)\xi,\;\;\;\mbox{where}\;\;\;\frac{\partial}{\partial t_j} = \frac{2E^j}{(m\lambda^2)^{j-1}} \frac{\partial}{\partial t_1}\;.
\end{equation}
\begin{figure}[h]
\begin{center}
\includegraphics[width=20pc]{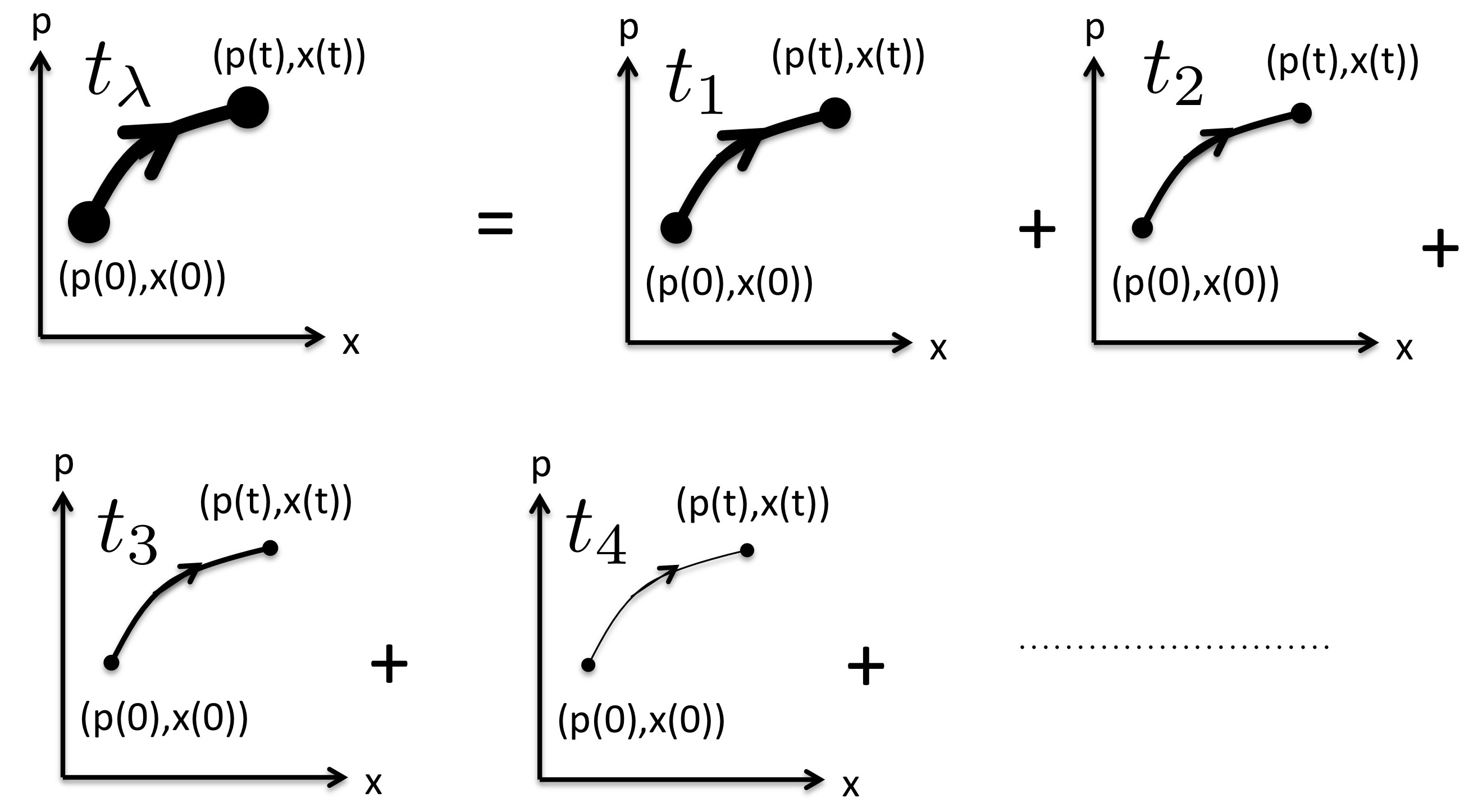}
\caption{\label{PS} The $\lambda$-flow associated with $H_{\lambda}$ on the phase space composes of infinite $j^{th}$ flows associated with $H_j$.}
\end{center}
\end{figure}\\
Equation (\ref{F1}) and (\ref{F2}) suggest us that the multiplicative Hamiltonian flow is constituted from infinite different Hamiltonian flows which they do describe the dynamics of the system on the same trajectory\footnote{Since all Hamiltonians produce the same equation of motion}. This means that if we work with the first Hamiltonian $H_1$, the time evolution of the system will be described by the standard Hamiltonian flow. If we work with the Hamiltonian $H_j$, when $j > 1$, the time evolution of the system will be described by the $j^{th}$ Hamiltonian flow which is at different rate scaling by the parameter $\lambda$. On the top level, if we work with $H_{\lambda}$ now the Hamiltonian flow is in the superimposition of infinite flows shown in figure \ref{PS}. In the case that  $\lambda$ goes to infinity, all Hamiltonian flows will be terminated, except for the standard one: $j=1$, which indeed agrees with the condition $\lim_{\lambda \to \infty} H_\lambda + m\lambda^2 = T + V$.

\section{The $\lambda$-extended class of canonical transformations}
\begin{figure}[h]
\begin{center}
\includegraphics[width=31pc, height=12pc]{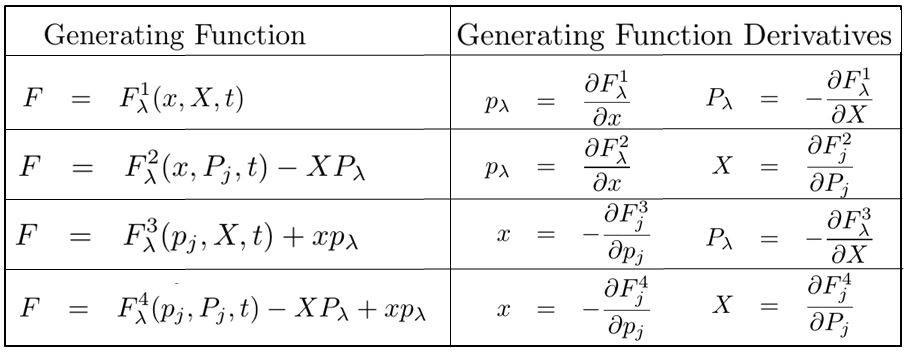}
\caption{\label{CT} Summary table for four types of $\lambda$-generating functions and $j = 1,2,3,...$}
\end{center}
\end{figure}

To derive four standard CTs, we start with the non-uniqueness property of the Lagrangian $L(\dot x,x) = L'(\dot X,X) + \frac{dF(X,x,t)}{ dt}$, where $L$ and $L'$ are in the additive form and $F$ is a generating function.  For the multiplicative form of Lagrangian and Hamiltonian, we may start with equation (\ref{Llam}) and apply the standard non-uniqueness property to $L_j$. We then obtain
\begin{equation}\label{FFF}
L_\lambda=\sum_{j=1}^{\infty}\frac{1}{j!}\Big(-\frac{1}{m\lambda^2}\Big)^{j-1}\Big(L'_j+\frac{dF_j}{dt}\Big) + m\lambda^2\;,\;\;\mbox{or}\;\;L_\lambda = L'_\lambda+\frac{dF_\lambda}{dt}\;,
\end{equation}
where
\begin{equation}\label{FF}
F_\lambda = \sum_{j=1}^{\infty}\frac{1}{j!}\left(-\frac{1}{m\lambda^2}\right)^{j-1}F_j = m\lambda^2\ln\left(1+\frac{F}{m\lambda^2}\right) \: , \: F_j = (j-1)! F^{j}\;,
\end{equation}
where $F_1=F$ is the standard generating function. The $\lambda$-generating function $F_\lambda$ is written in the form of infinite series of $F_j$ given in eq. (\ref{FF}) and $\lim_{\lambda\rightarrow\infty}F_\lambda=F$. Using eq. (\ref{FFF}), we then derive new four types of CTs called the $\lambda$-extended CTs mapping $(p_\lambda,x)$ to $(P_\lambda,X)$ shown in figure \ref{CT}.
\\
Each type of CTs can be considered as a generator for infinite hierarchy by expanding the equations with respect with the parameter $\lambda$. Obviously, all the cases in the table will be reduced to the standard CTs when $\lambda$ go to infinity.

\section{Conclusion}
We succeeded to construct the infinite hierarchy Legendre transformations to connect Hamiltonian hierarchy $\{H_1,H_2,... \}$ with Lagrangian hierarchy $\{ L_1,L_2,...\}$ for a system with 1 degree of freedom. For the geometric structure of the phase space, the multiplicative Hamiltonian flow came from infinite hierarchy flows associated with infinite Hamiltonian hierarchy. The role of the parameter $\lambda$ was the flow scaling in the Hamiltonian dynamics. New four types of canonical transformations were successfully derived and infinite set for each type was discovered. The result in this study is totally new and provides a new feature of classical dynamics for any system with 1 degree of freedom. This suggests us that actually nature may hide us an intriguing mathematical structure and was not so obvious to discover. 
\\\\
Finally, we will leave some remarks for future investigations. The first one is that it might be interesting to extend the idea to case of many degrees of freedom and to see whether there are any new features both in physics and mathematics point of views. Another point is that the role of the parameter $\lambda$ in the Lagrangian dynamics on configuration space is not yet clear at the moment. Then it would be good if we could understand it in the near future in order to have a complete picture.

\ack
Saksilpa Srisukson was supported by Junior Science Talent Project, a project of National Science and Technology Development Agency (NSTDA). Sikarin Yoo-Kong was supported by the Theoretical and Computational Science (TaCS) Center under ComputationaL and Applied Science for Smart Innovation Cluster (CLASSIC), Faculty of Science, KMUTT.


\section*{Reference}
\begin{thebibliography}{9}
\bibitem{Su} 
Surawuttinack, K., Yoo-Kong, S., Tanasittikosol, M, 2016, Multiplicative form of the Lagrangian, Mathematical and Theoretical Physics, \textbf{189}(3), pp:1693-1711.
\end{thebibliography}
\end{document}